\documentclass[authoryear,preprint,review,12pt]{elsarticle}

\usepackage{amssymb}
\usepackage{epsfig}

\newtheorem{theorem}{Theorem}
\newtheorem{corollary}{Corollary}
\newtheorem{lemma}{Lemma}

\newtheorem{proposition}{Proposition}

\newcommand{\indep}{\;\, \rule[0em]{.03em}{.67em} \hspace{-.25em}
\rule[0em]{.65em}{.03em} \hspace{-.25em} \rule[0em]{.03em}{.67em}\;\,}

\begin{document}

\begin{frontmatter}
\title{Sliced Inverse Moment Regression Using Weighted Chi-Squared Tests for Dimension Reduction}
\author{Zhishen Ye\fnref{label1}}
\author{Jie Yang\corref{cor1}\fnref{label2,label3}}
\address[label1]{Amgen Inc., Thousand Oaks, CA 91320-1799, USA}
\address[label2]{Department of Mathematics, Statistics, and Computer Science, University of Illinois at Chicago, Chicago, IL 60607-7045, USA}
\fntext[label3]{The authors thank Robert Weiss for comments on an earlier draft.
                They are also very grateful to Bing Li and Shaoli Wang for sharing their computer program.}
\cortext[cor1]{Corresponding author at: Department of Mathematics, Statistics, and Computer Science (MC 249), University of Illinois at Chicago,
               851 South Morgan Street, SEO 322, Chicago, Illinois 60607, USA.
               Tel.:+13124133748; fax:+13129961491.
               E-mail address: jyang06@math.uic.edu (J. Yang).}

\begin{abstract}
We propose a new method for dimension reduction in regression using the first
two inverse moments. We develop corresponding weighted chi-squared tests for
the dimension of the regression. The proposed method considers linear combinations
of Sliced Inverse Regression (SIR) and the method using a new candidate matrix
which is designed to recover the entire inverse second moment subspace. The
optimal combination may be selected based on the p-values derived from the dimension tests.
Theoretically, the proposed method, as well as Sliced Average Variance Estimate (SAVE),
are more capable of recovering the complete central dimension reduction subspace
than SIR and Principle Hessian Directions (pHd). Therefore it can substitute for
SIR, pHd, SAVE, or any linear combination of them at a theoretical level. Simulation
study indicates that the proposed method may have consistently greater power than SIR,
pHd, and SAVE.
\end{abstract}


\end{frontmatter}

\section{Introduction}

The purpose of the regression of a univariate response $y$ on a $p$-dimensional
predictor vector ${\mathbf{x}}$ is to make inference on the conditional distribution of $y|{\mathbf{x}}$.
Following \citet{cook1998b}, ${\mathbf{x}}$ can be replaced by its standardized version
\begin{eqnarray}
{\mathbf{z}}= [\Sigma_{{\mathbf{x}}}]^{-{1}/{2}} ({\mathbf{x}}-\mu_{{\mathbf{x}}})\ ,
\label{standardize}
\end{eqnarray}
where $\mu_{{\mathbf{x}}}$ and $\Sigma_{{\mathbf{x}}}$ denote
the mean and covariance matrix of ${\mathbf{x}}$ respectively assuming non-singularity
of $\Sigma_{{\mathbf{x}}}$.

The goal of {\it dimension reduction in regression} is to
find out a $p\times d$ matrix $\gamma$ such that
\begin{eqnarray}
y \indep {\mathbf{z}} | \gamma' {\mathbf{z}}\ ,
\label{modelcdrs}
\end{eqnarray}
where ``$\indep$" indicates independence.
Then the $p$-dimensional {${\mathbf{z}}$} can be replaced
by the $d$-dimensional vector $\gamma' {\mathbf{z}}$
without specifying any parametric model
and without losing any information on predicting $y$.
The column space ${\rm Span}\{\gamma\}$ is called a
{\it{dimension reduction subspace}}.
The smallest applicable $d$ is called the {\it dimension of the regression}.

Based on the inverse mean ${\rm E}({\mathbf{z}}|y)$, \citet{li1991a} proposed Sliced Inverse
Regression (SIR) for dimension reduction in regression.
It is realized that SIR can not recover the symmetric dependency \citep*{li1991b,cook1991}.
After SIR, many dimension reduction methods have been introduced.
Sliced Average Variance Estimate (SAVE) proposed by \citet*{cook1991}
and Principle Hessian Directions (pHd) proposed by \citet*{li1992} are another two popular ones.
Both pHd and SAVE refer to the second inverse moment, centered or non-centered.
Compared with SAVE, pHd can not detect certain dependency hidden in the second moment
\citep*{yin2002,ye2003} and the linear dependency \citep*{li1992,cook1998a}.
Among those dimension reduction methods using only the first two inverse moments,
SAVE seems to be the preferred one.  Nevertheless, SAVE is not always the winner.
For example, \citet*{ye2003} implied that
a linear combination of SIR and pHd may perform better than SAVE
in some cases.  It is not surprising since \citet{li1991b} already suggested that
a suitable combination of two different methods might sharpen the dimension reduction results.
\citet*{ye2003}  further proposed that
a bootstrap method could be used to pick up the ``best" linear combination of two
known methods, as well as the dimension of the regression,
in the sense of the variability of the estimators,
although lower variability under the bootstrap procedure does not
necessarily lead to a better estimator. \citet*{liwang2007} pointed out
that linear combinations of two known methods selected by the bootstrap criterion
may not perform as well as a single new method, their Directional
Regression method (DR), even though the bootstrap one is computationally intensive.

This article aims to develop a new class of, instead of a single one,
dimension reduction methods using only the first two inverse moments,
as well as the corresponding large sample tests for the dimension of the regression
and an efficient criterion for selecting a suitable candidate from the class.
Theoretically, it can cover SIR, pHd, SAVE and their linear combinations.
Practically, it can achieve higher power in recovering the dimension reduction subspace.
In Section~2, we review the necessary dimension reduction context.
In Section~3, we introduce a simple candidate matrix $M_{{\mathbf{z}}{\mathbf{z}}'|y}$
which targets the entire inverse second moment subspace.
It is indeed the candidate matrix of an intermediate method
between pHd and SAVE.
In Section~4, we propose a new class of dimension reduction methods
called Sliced Inverse Moment Regression (SIMR),
along with weighted chi-squared tests for
the dimension of the regression.
In Section~5, we use SIMR to analyze a simulated example
and illustrate how to select a good candidate of SIMR.
Simulation study shows that SIMR may have
consistently greater power than SIR, pHd, and SAVE, as well as DR
and another new method Inverse Regression Estimator \citep*{cookni2005}.
In Section~\ref{sectionozonedata}, a real example is used to
illustrate how the proposed method works.
It is implied that a class of dimension reduction methods, along with
a suitable criterion for choosing
a good one among them, may be preferable in practice to any single method.
We conclude this article with discussion and proofs of the results presented.

\section{Dimension Reduction Context}

\subsection{Central Dimension Reduction Subspace (CDRS)}

\citet*{cook1994b,cook1996} introduced the notion of
{\it{central dimension reduction subspace}} (CDRS),
denoted by $S_{y|{\mathbf{z}}}$, which is the intersection of all dimension reduction subspaces.
Under fairly weak restrictions, the CDRS $S_{y|{\mathbf{z}}}$ is still a dimension reduction subspace.

In this article, we always assume that $S_{y|{\mathbf{z}}}$ is a dimension reduction subspace
and that the columns of $\gamma$ is an orthonormal basis of $S_{y|{\mathbf{z}}}$.
In practice,
we usually first transform the original data $\{{\mathbf{x}}_i\}$ into their standardized version
$\{{\mathbf{z}}_i\}$ by replacing $\Sigma_{\mathbf{x}}$ and $\mu_{\mathbf{x}}$  in (\ref{standardize})
with their usual sample estimates $\hat{\Sigma}_{{\mathbf{x}}}$ and $\hat{\mu}_{{\mathbf{x}}}$.
Then we can estimate $S_{y|{\mathbf{x}}}$ by
\begin{eqnarray*}
\hat{S}_{y|{\mathbf{x}}}=[\hat{\Sigma}_{{\mathbf{x}}}]^{-{1}/{2}} \hat{S}_{y|{\mathbf{z}}}\ ,
\end{eqnarray*}
where $\hat{S}_{y|{\mathbf{z}}}$ is an estimate of ${S}_{y|{\mathbf{z}}}$.
Therefore, the goal of dimension reduction in regression is to
find out the dimension of the regression $d$ and the CDRS
${S}_{y|{\mathbf{z}}}={\rm Span}\{\gamma\}$.

Following \citet*{li1991a} and \citet*{cook1998b}, we also assume:
(1) $E({\mathbf{z}}|\gamma' {\mathbf{z}}) = P_{\gamma}{\mathbf{z}}$,
where $P_{\gamma}=\gamma \gamma'$, known as the {\it linearity condition};
(2) ${\rm Var} ({\mathbf{z}}|\gamma' {\mathbf{z}}) = Q_{\gamma}$,
where $Q_{\gamma}={\rm I} - P_{\gamma}$, known as the
{\it constant covariance condition}.
These two conditions hold if ${\mathbf{z}}$ is normally distributed,
although the normality is not necessary.

\subsection{Candidate Matrix}
\label{sectioncandidatematrix}

\citet*{ye2003} introduced the concept of {\it candidate matrix}, which is a
$p\times p$ matrix $A$ satisfying $A = P_{\gamma} A P_{\gamma}$.
They showed that any eigenvector corresponding to any nonzero eigenvalue of $A$ belongs to
the CDRS ${\rm Span}\{\gamma\}$.
Besides, the set of all candidate matrices, denoted by $\mathcal{M}$,
is closed under
scalar multiplication, transpose, addition, multiplication, and
thus under linear combination and expectation.


They also showed that the matrices $[\mu_1(y) \mu_1(y)']$ and $[\mu_2(y)-{\rm I}]$ belong to
$\mathcal{M}$ for all $y$, where $\mu_1(y)={\rm E}({\mathbf{z}}|y)$ and $\mu_2(y)={\rm E}({\mathbf{z}}{\mathbf{z}}'|y)$.
They proved that the symmetric matrices that SIR, SAVE, and $y$-pHd estimate all belong to $\mathcal{M}$:
\begin{eqnarray*}
M_{{\rm SIR}} &=& {\rm Var} ({\rm E}({\mathbf{z}}|y)) = {\rm E} [\mu_1(y) \mu_1(y)']\ , \\
M_{{\rm SAVE}} &=& {\rm E}[(I-{\rm Var} ({\mathbf{z}}|y))^2] \\
 &=& {\rm E} ([\mu_1(y) \mu_1(y)']^2 + [\mu_2(y) - {\rm I}]^2  \\
 & &  -  [\mu_1(y) \mu_1(y)'][\mu_2(y) - {\rm I}] -  [\mu_2(y) - {\rm I}][\mu_1(y) \mu_1(y)'] )\ , \\
M_{y{\rm -pHd}} &=& {\rm E}[(y-{\rm E}(y)){\mathbf{z}}{\mathbf{z}}'] = {\rm E}[y (\mu_2(y)-{\rm I})]\ .
\end{eqnarray*}

\section{Candidate Matrix $M_{{\mathbf{z}}{\mathbf{z}}'|y}$}


\subsection{A Simple Candidate Matrix}

The matrices $[\mu_1(y)\mu_1(y)']$ and $[\mu_2(y)-{\rm I}]$ are actually two fundamental components of
$M_{{\rm SIR}}$, $M_{{\rm SAVE}}$, and $M_{y{\rm -pHd}}$ (see Section~\ref{sectioncandidatematrix}).
$M_{{\rm SIR}}$ only involves the first component $[\mu_1(y)\mu_1(y)']$,
while both $M_{{\rm SAVE}}$ and $M_{y{\rm -pHd}}$ share the second component $[\mu_2(y)-{\rm I}]$.
Realizing that this common feature may lead to the connection between SAVE and pHd,
we investigate the behavior of the matrix $[\mu_2(y)-{\rm I}]$.
To avoid the inconvenience due to ${\rm E} ([\mu_2(y)-{\rm I}])=0$, we define
\begin{eqnarray*}
M_{{\mathbf{z}}{\mathbf{z}}'|y} &=& {\rm E} ( [ {\rm E}({\mathbf{z}}{\mathbf{z}}'-{\rm I}|y) ]^2 ) = {\rm E} ([\mu_2(y)-{\rm I}]^2).
\end{eqnarray*}
Note that $M_{{\mathbf{z}}{\mathbf{z}}'|y}$ takes a simpler form
than the rescaled version of $\mbox{sirII}$ \citep[Remark R.3]{li1991b}
while still keeping the theoretical comprehensiveness.
It also appears as a component in one expression of the {\it directional regression} matrix $G$
\citep*[eq.(4)]{liwang2007}.
We choose its form as simple as possible for less complicated large sample
test and potentially greater test power.
To establish the relationship between $M_{y{\rm -pHd}}$ and $M_{{\mathbf{z}}{\mathbf{z}}'|y}$,
we need:
\begin{lemma}
Let $M$ be a $p\times q$ random matrix defined on a probability space
$(\Omega,{\cal F},P)$, then there exists an event $\Omega_0 \in {\cal F}$ with probability $1$,
such that,
$${\rm Span} \{ E ( MM' ) \}
    = {\rm Span} \{ M(\omega), \omega \in \Omega_0 \}.$$
\label{asaveone}
\end{lemma}
A similar result can also be found in \citet*[Proposition 2(i)]{yin2003}.
The lemma here is more general.  By
the definition of $M_{{\mathbf{z}}{\mathbf{z}}'|y}$,
\begin{corollary}
${\rm Span} \{ M_{{\mathbf{z}}{\mathbf{z}}'|y} \} = {\rm Span} \{ [\mu_2(y)-{\rm I}], y \in \Omega(y)\}$,
where $\Omega(y)$ is the support of $y$.
\label{asavetwo}
\end{corollary}
Based on Corollary \ref{asavetwo}, \citet*[Lemma 3]{ye2003},
and the fact that $[\mu_2(y)-{\rm I}] \in \mathcal{M}$ for all $y$,
matrix $M_{{\mathbf{z}}{\mathbf{z}}'|y}$ is in fact a candidate matrix too.
Corollary \ref{asavetwo} also implies a strong connection between $M_{y{\rm -pHd}}$
and $M_{{\mathbf{z}}{\mathbf{z}}'|y}$:
\begin{corollary}
$ {\rm Span}\{ M_{y{\rm -pHd}} \} \subseteq {\rm Span}\{ M_{{\mathbf{z}}{\mathbf{z}}'|y} \} $.
\label{asavethree}
\end{corollary}

To further understand the relationship between
$M_{y{\rm -pHd}}$ and $M_{{\mathbf{z}}{\mathbf{z}}'|y}$,
recall the {\it{central $k$-th moment dimension reduction subspace}} \citep*{yin2003},
$S_{y|{\mathbf{z}}}^{(k)}={\rm Span}\{\eta^{(k)}\}$.
The corresponding random vector $(\eta^{(k)})' {\mathbf{z}}$
contains all the available information about $y$ from the first $k$
conditional moments of $y|{\mathbf{z}}$.  In other words,
$\left.y \indep \{{\rm E}(y|{\mathbf{z}}),\ldots,{\rm E}(y^k|{\mathbf{z}})\} \right| (\eta^{(k)})' {\mathbf{z}}$.
Similar to
\[
{\rm Span}\{{\rm E}(y{\mathbf{z}}),\ldots,{\rm E}(y^k{\mathbf{z}})\}
= {\rm Span}\{{\rm E}(y \mu_1(y)),\ldots,{\rm E}(y^k \mu_1(y))\}
\subseteq S_{y|{\mathbf{z}}}^{(k)} \subseteq S_{y|{\mathbf{z}}},
\]
the subspace ${\rm Span}\{ {\rm E}(y [\mu_2(y)-{\rm I}]),\ldots,{\rm E}(y^k [\mu_2(y)-{\rm I}])\}$ is also contained in
$S_{y|{\mathbf{z}}}^{(k)}$.
Parallel to \citet*[Proposition 4]{yin2002}, the result on $M_{{\mathbf{z}}{\mathbf{z}}'|y}$ is:
\begin{proposition}
(a) If $y$ has finite support $\Omega(y)=\{a_0,\ldots,a_k\}$, then
\begin{eqnarray*}
{\rm Span} \{ M_{{\mathbf{z}}{\mathbf{z}}'|y} \}  = {\rm Span} \{ {\rm E}[y^i (\mu_2(y)-{\rm I})],\ i=1,\ldots,k \}.
\end{eqnarray*}
(b) If $y$ is continuous and $\mu_2(y)$ is continuous on $y$'s support $\Omega(y)$,
then
\begin{eqnarray*}
{\rm Span} \{ M_{{\mathbf{z}}{\mathbf{z}}'|y} \}  = {\rm Span} \{ {\rm E}[y^i (\mu_2(y)-{\rm I})],\ i=1,2,\ldots \}.
\end{eqnarray*}
\label{asavefour}
\end{proposition}
According to Proposition~\ref{asavefour} and \citet*[Proposition 4]{yin2002},
the relationship between
${\rm E}[y (\mu_2(y)-{\rm I})]=M_{y{\rm -pHd}}$ and $M_{{\mathbf{z}}{\mathbf{z}}'|y}$
is fairly comparable with the relationship between
${\rm E}(y \mu_1(y) ) = {\rm E}(y{\mathbf{z}})$ and $M_{{\rm SIR}}$.
Both ${\rm E}(y{\mathbf{z}})$ and $M_{y{\rm -pHd}}$ actually target
the central mean (first moment) dimension reduction subspace \citep*{cook2002},
while $M_{{\rm SIR}}$ and $M_{{\mathbf{z}}{\mathbf{z}}'|y}$ target the central $k$-th moment dimension
reduction subspace given any $k$, or equivalently the CDRS $S_{y|{\mathbf{z}}}$ as $k$ goes to infinite.
In order to understand the similarity from another perspective,
recall the {\it inverse mean subspace} of $S_{y|{\mathbf{z}}}$ \citep*{yin2002}:
$$S_{{\rm E}({\mathbf{z}}|y)}={\rm Span}\{{\rm E}({\mathbf{z}}|y), y\in \Omega(y)\}.$$
Similarly, we define the {\it inverse second moment subspace} of $S_{y|{\mathbf{z}}}$:
$${\rm Span}\{{\rm E}({\mathbf{z}}{\mathbf{z}}'|y)-{\rm I}, y\in \Omega(y)\}.$$
By definition, matrices $M_{{\rm SIR}}$ and $M_{{\mathbf{z}}{\mathbf{z}}'|y}$ are designed to recover
the entire inverse mean subspace and the entire inverse second moment subspace respectively,
while ${\rm E}(y{\mathbf{z}})$ and $M_{y{\rm -pHd}}$ are only able to recover portions of those subspaces.
We are therefore interested in combining matrices $M_{{\rm SIR}}$ and $M_{{\mathbf{z}}{\mathbf{z}}'|y}$
because they are both comprehensive.

\subsection{ SAVE versus SIR and pHd}

\citet*{ye2003} showed that
\begin{eqnarray}
{\rm Span}\{  M_{{\rm SIR}} \} \subseteq {\rm Span}\{  M_{{\rm SAVE}} \},
\label{yeweiss}
\end{eqnarray}
We then prove further the following proposition:
\begin{proposition}
${\rm Span}\{ M_{{\rm SAVE}} \} = {\rm Span}\{ M_{{\rm SIR}} \} + {\rm Span}\{ M_{{\mathbf{z}}{\mathbf{z}}'|y} \}.$
\label{asavefive}
\end{proposition}
A straightforward result following Proposition \ref{asavefive} and Corollary \ref{asavethree} is:
\begin{corollary}
$ {\rm Span}\{ M_{y{\rm -pHd}} \}, {\rm Span}\{ M_{{\rm SIR}} \}, {\rm Span}\{ M_{{\mathbf{z}}{\mathbf{z}}'|y} \}
    \subseteq {\rm Span} \{ M_{{\rm SAVE}} \}. $
\label{asavesix}
\end{corollary}
Corollary \ref{asavesix}
explains why SAVE is able to provide better estimates of the CDRS
than SIR and $y$-pHd in many cases.

\section{Sliced Inverse Moment Regression Using
    Weighted Chi-Squared Tests}

\subsection{Sliced Inverse Moment Regression}

In order to simplify the candidate matrices using the first two inverse moments
and still keep the comprehensiveness of SAVE,
a natural idea is to combine $M_{{\mathbf{z}}{\mathbf{z}}'|y}$ with $M_{{\rm SIR}}$ as follows:
\[
\alpha M_{{\rm SIR}} + (1-\alpha) M_{{\mathbf{z}}{\mathbf{z}}'|y}
= {\rm E} (\alpha[\mu_1(y) \mu_1(y)'] +  (1-\alpha)[\mu_2(y) - {\rm I}]^2 ),
\]
where $\alpha\in (0,1)$. We call this matrix $M^{(\alpha)}_{{\rm SIMR}}$
and the corresponding dimension reduction method {\it Sliced Inverse Moment Regression} (SIMR or SIMR$_{\alpha}$).
Note that the combination here is simpler than the $\mbox{SIR}_\alpha$ method \citep*{li1991b,gannoun2003}
while retaining the least requirement on comprehensiveness.
Actually, for any $\alpha\in (0,1)$, SIMR$_\alpha$ is as comprehensive as SAVE at a theoretical level based on
the following proposition:
\begin{proposition}
${\rm Span}\{M^{(\alpha)}_{{\rm SIMR}}\}={\rm Span}\{M_{{\rm SAVE}}\},\ \forall \alpha \in (0,1).$
\label{asaveseven}
\end{proposition}
Combined with Corollary~\ref{asavesix}, we know that any linear combination of SIR, pHd and
SAVE can be covered by SIMR$_\alpha$:
\begin{corollary}
${\rm Span}\{a M_{{\rm SIR}}+ b M_{y{\rm -pHd}} + c M_{{\rm SAVE}}\}\subseteq {\rm Span}\{M^{(\alpha)}_{{\rm SIMR}}\}$,
where $a, b$, and $c$ are arbitrary real numbers.
\label{simr8}
\end{corollary}
Note that the way of constructing SIMR$_\alpha$ makes it easier to develop a
corresponding large sample test for the dimension of the regression (Section~4.3).

From now on, we assume that the data $\{(y_i,{\mathbf{x}}_i)\}_{i=1,\ldots,n}$ are i.i.d.~from
a population which has finite first four moments and conditional moments.

\subsection{Algorithm for SIMR$_\alpha$}

Given i.i.d.~sample $(y_1,{\mathbf{x}}_1)$,...,$(y_n,{\mathbf{x}}_n)$,
first standardize ${\mathbf{x}}_i$ into $\hat{\mathbf{z}}_i$,
sort the data by $y$,
and divide the data into $H$ slices with intraslice sample sizes
$n_h$, $h=1,\ldots,H$.
Secondly construct the intraslice sample means
$\overline{({\mathbf{z}}{\mathbf{z}}')}_h$ and $\bar{{\mathbf{z}}}_h$:
\begin{eqnarray*}
\overline{({\mathbf{z}}{\mathbf{z}}')}_h &=& \frac{1}{n_h}\sum_{i=1}^{n_h} \hat{\mathbf{z}}_{ih} \hat{\mathbf{z}}_{ih}'\ ,
\\
\bar{{\mathbf{z}}}_h &=& \frac{1}{n_h}\sum_{i=1}^{n_h} \hat{\mathbf{z}}_{ih}\ ,
\end{eqnarray*}
where $\hat{\mathbf{z}}_{ih}$'s are predictors falling into slice $h$.
Thirdly calculate
\begin{eqnarray*}
\hat{M}^{(\alpha)}_{{\rm SIMR}} &=& \sum_{h=1}^H  \hat{f}_h
 \left( (1-\alpha)[\overline{({\mathbf{z}}{\mathbf{z}}')}_h - {\rm I}_p] [\overline{({\mathbf{z}}{\mathbf{z}}')}_h - {\rm I}_p]'
 + \alpha [\bar{{\mathbf{z}}}_h] [\bar{{\mathbf{z}}}_h]'\right)
\\
&=& \hat{U}_{n} \hat{U}_{n}'\ ,
\end{eqnarray*}
where $\hat{f}_h = n_h/n$ and
\begin{eqnarray*}
\hat{U}_{n} =   \left(
    \ldots, \sqrt{1-\alpha}\ [\overline{({\mathbf{z}}{\mathbf{z}}')}_h - {\rm I}_p]\sqrt{\hat{f}_h}, \ldots ,
\,
        \ldots,\sqrt{\alpha}\ \bar{{\mathbf{z}}}_h \sqrt{\hat{f}_h} ,\ldots
        \right)_{p \times (pH+H)}.
\end{eqnarray*}
Finally calculate the eigenvalues
$\hat{\lambda}_{1} \geq \cdots \geq \hat{\lambda}_{p}$ of $\hat{M}^{(\alpha)}_{{\rm SIMR}}$
and the corresponding eigenvectors $\hat{\gamma}_1, \ldots, \hat{\gamma}_p$~.
Then ${\rm Span}\{ \hat{\gamma}_{1},\ldots,\hat{\gamma}_{d} \}$ is an estimate of the
CDRS ${\rm Span}\{\gamma\}$, where $d$ is determined by the weighted chi-squared
test described in the next section.

\subsection{A Weighted Chi-Squared Test for SIMR$_\alpha$}
\label{weightedtest}

Define the population version of $\hat{U}_{n}$:
\begin{eqnarray}
& & B \nonumber\\
&=& \left(
    \ldots ,
    \sqrt{1-\alpha}\ [{\rm E}({\mathbf{z}}{\mathbf{z}}'|\tilde{y}=h) - {\rm I}_p]\sqrt{f_h},
    \ldots,
    \sqrt{\alpha}{\rm E}({\mathbf{z}}|\tilde{y}=h)\sqrt{f_h},
    \ldots
      \right)
\nonumber\\
&=&\left( (\Gamma_{11})_{p \times d} , (\Gamma_{12})_{p \times (p-d)} \right)
   \left( \begin{array}{cc} D_{d \times d} & 0 \\ 0 & 0 \end{array} \right)
   \left( \begin{array}{l} (\Gamma_{21}')_{d \times (pH+H)}
        \\ (\Gamma_{22}')_{(pH+H-d) \times (pH+H)} \end{array} \right)
\label{svd}
\end{eqnarray}
where $\tilde{y}$ is a slice indicator with
$\tilde{y} \equiv h$ for all observations falling into slice $h$,
${f_h} = P( \tilde{y} = h )$ is the population version of $\hat{f}_h$,
and (\ref{svd}) is the {\it singular value decomposition} of $B$.

Denote $\tilde{U}_{n}=\sqrt{n}(\hat{U}_{n} - B)$.
By the multivariate central limit theorem and the multivariate version of Slutsky's theorem,
$\tilde{U}_n$ converges in distribution to a certain random $p \times (pH+H)$ matrix $U$
as $n$ goes to infinity \citep*{gannoun2003}.
Note that the singular values are invariant under right and left multiplication by orthogonal matrices.
Based on \citet*[Theorem 4.1 and 4.2]{eaton1994},
the asymptotic distribution of
the smallest $(p-d)$ singular values of $\sqrt{n} \hat{U}_n$ is
the same as the asymptotic distribution of
the corresponding singular values of the following $(p-d) \times (pH+H-d)$ matrix:
\begin{eqnarray}
\sqrt{n} \Gamma_{12}' \hat{U}_n \Gamma_{22}.
\label{eatontylerone}
\end{eqnarray}
Construct statistic
\begin{eqnarray*}
\hat{\Lambda}_{d} = n  \sum_{h=d+1}^p \hat{\lambda}_h,
\end{eqnarray*}
which is the sum of the squared smallest $(p-d)$
singular values of $\sqrt{n} \hat{U}_n$.
Then the asymptotic distribution of $\hat{\Lambda}_{d}$ is the same as
that of the sum of the squared singular values of (\ref{eatontylerone}).  That is
\[
n {\rm Trace}(  [\Gamma_{12}' \hat{U}_n \Gamma_{22}] [\Gamma_{12}' \hat{U}_n \Gamma_{22}]' )
= n [{\rm Vec} (\Gamma_{12}' \hat{U}_n \Gamma_{22})]'
    [{\rm Vec} (\Gamma_{12}' \hat{U}_n \Gamma_{22})],
\]
where ${\rm Vec}(A_{r \times c})$ denotes
$({{a_1}}',\ldots,{{a_c}}')'_{rc \times 1}$ for
any matrix $A=({{a_1}},\ldots,{{a_c}})$.
By central limit theorem and Slutsky's theorem again,
\begin{eqnarray*}
{\rm Vec}(\tilde{U}_n) \stackrel{\mathcal{L}}{\rightarrow} N_{(p^2H+pH)}(0,V)
\end{eqnarray*}
for some nonrandom $(p^2H+pH) \times (p^2H+pH)$ matrix $V$.
Thus,
\begin{eqnarray*}
\sqrt{n} [{\rm Vec} (\Gamma_{12}' \hat{U}_n \Gamma_{22})]  \stackrel{\mathcal{L}}{\rightarrow}
 N_{(p-d)(pH+H-d)} (0, W),
\end{eqnarray*}
where $W=[\Gamma_{22}' \otimes \Gamma_{12}'] V [\Gamma_{22}' \otimes \Gamma_{12}']'$
is a $(p-d)(pH+H-d) \times (p-d)(pH+H-d)$ matrix.
Combined with Slutsky's theorem, it yields the following theorem:
\begin{theorem}
The asymptotic distribution of $\hat{\Lambda}_{d}$ is the same as that of
\begin{eqnarray*}
\sum_{i=1}^{(p-d)(pH+H-d)} \alpha_i K_i
\end{eqnarray*}
where the $K_i$'s are independent $\chi_1^2$ random variables,
and $\alpha_i$'s are the eigenvalues of the matrix $W$.
\label{theoremasaveone}
\end{theorem}

Clearly,
a consistent estimate of $W$ is needed
for testing the dimension of the regression based on
Theorem \ref{theoremasaveone}.
The way we define $M^{(\alpha)}_{{\rm SIMR}}$
allows us to partition $\hat{U}_n$ into
\begin{eqnarray*}
\hat{U}_{n,1} &=&   \left(
    \ldots, \sqrt{1-\alpha}\ [\overline{({\mathbf{z}}{\mathbf{z}}')}_h - {\rm I}_p]\sqrt{\hat{f}_h}, \ldots ,
            \right)_{p \times pH}\ ,
\\
\hat{U}_{n,2} &=&   \left(
    \ldots,\sqrt{\alpha}\ \bar{{\mathbf{z}}}_h \sqrt{\hat{f}_h} ,\ldots
        \right)_{p \times H}\ .
\end{eqnarray*}
The asymptotic distribution of the matrix $\hat{U}_{n,2}$ has been fully
explored by \citet*{bura2001}, resulting in a weighted chi-squared test
for SIR.  The similar techniques can also be applied on the matrix
$\hat{U}_{n,1}$, and therefore the matrix  $\hat{U}_n$ as a whole, although
the details are much more complicated.

Define the population versions of $\hat{U}_{n,1}$ and $\hat{U}_{n,2}$,
\begin{eqnarray*}
B_{1} &=& \left(
    \ldots ,
   \sqrt{1-\alpha}\ [{\rm E}({\mathbf{z}}{\mathbf{z}}'|\tilde{y}=h) - {\rm I}_p]\sqrt{f_h},
    \ldots
      \right)_{p \times pH},
\\
B_{2} &=& \left(
    \ldots,
    \sqrt{\alpha}\ {\rm E}({\mathbf{z}}|\tilde{y}=h)\sqrt{f_h},
    \ldots
      \right)_{p \times H}.
\end{eqnarray*}
Then
$\hat{U}_n = \left(
    \hat{U}_{n,1} , \hat{U}_{n,2}
      \right)$,
and $B = \left(
    B_1,B_2
      \right)$.

Let $f$, $\hat{f}$ and $1_H$ be $H \times 1$ vectors with
elements $f_h$, $\hat{f}_h$ and $1$ respectively;
let $G$ and $\hat{G}$ be $H \times H$ diagonal matrices with
diagonal entries $\sqrt{{f}_h}$ and $\sqrt{\hat{f}_h}$ respectively;
and let
\[
\hat{F} = ({\rm I}_H-\hat{f} 1_H'),\>\>
{F} = ({\rm I}_H-{f} 1_H'),
\]
\[
\left( \begin{array}{l} (\Gamma_{21}') \\ (\Gamma_{22}') \end{array} \right)
=
\left( \begin{array}{ll}   (\Gamma_{211}')_{d \times pH}        & (\Gamma_{212}')_{d \times H}
            \\ (\Gamma_{221}')_{(pH+H-d) \times pH}     & (\Gamma_{222}')_{(pH+H-d) \times H}
\end{array} \right).
\]
Finally, define four matrices
\begin{eqnarray*}
M &=& (\dots, {\rm E}({\mathbf{x}}|\tilde{y}=h),\ldots)_{p \times H},
\\
N &=& (\dots, {\rm E}({\mathbf{x}}'|\tilde{y}=h),\ldots)_{1 \times pH} = {\rm Vec}(M)',
\\
O &=& (\dots, {\rm E}({\mathbf{x}}{\mathbf{x}}'|\tilde{y}=h),\ldots)_{p \times pH},
\\
C &=& [ O - M ({\rm I}_H \otimes {\mu}_{{\mathbf{x}}}') - {\mu}_{{\mathbf{x}}} N]_{p \times pH},
\end{eqnarray*}
and their corresponding sample versions $M_n$, $N_n$, $O_n$, and $C_n$.
By the central limit theorem,
\begin{eqnarray*}
\sqrt{n}  {\rm Vec} ( [(C_n,M_n) - (C,M)])
    \stackrel{\mathcal{L}}{\rightarrow}
N_{(p^2H+pH)}(0,\Delta)
\end{eqnarray*}
for a nonrandom $(p^2H+pH) \times (p^2H+pH)$ matrix $\Delta$.
As a result,
\begin{theorem}The covariance matrix in Theorem~\ref{theoremasaveone} is
\[
W =   (K
    \Gamma_{22}
    )'
    \otimes
    (\Gamma_{12}' {\Sigma}_{{\mathbf{x}}}^{-1/2})
\Delta
 (K    \Gamma_{22}
    )
    \otimes
    (\Gamma_{12}' {\Sigma}_{{\mathbf{x}}}^{-1/2} )'\ ,
\]
where
\[
K=    \left( \begin{array}{cc}
        \sqrt{1-\alpha} ({F} {G})  \otimes {\Sigma}_{{\mathbf{x}}}^{-1/2} & 0
        \\
        0 & \sqrt{\alpha}\ F G
    \end{array}
    \right)
\]
\label{theoremasavetwo}
\end{theorem}

The only difficulty left now is to obtain a consistent estimate of $\Delta$.
By the central limit theorem,
\begin{eqnarray*}
\sqrt{n}  {\rm Vec} ( [(O_n,M_n,\hat{\mu}_{{\mathbf{x}}}) - (O,M,\mu_{{\mathbf{x}}})] )
    \stackrel{\mathcal{L}}{\rightarrow}
N_{(p^2H+pH+p)}(0,\Delta_0)
\end{eqnarray*}
where $\Delta_0$ is a nonrandom $(p^2H+pH+p) \times (p^2H+pH+p)$ matrix,
with details shown in the Appendix.  On the other hand,
\begin{eqnarray*}
{\rm Vec} (C_n,M_n) &=&
    \left( \begin{array}{ccc}
        {\rm I}_{p^2H} &     -{\rm I}_H \otimes \hat{\mu}_{{\mathbf{x}}} \otimes {\rm I}_p-{\rm I}_{pH}\otimes\hat{\mu}_{{\mathbf{x}}}
                        & 0
        \\
        0 &         {\rm I}_{pH}                     & 0
    \end{array}
    \right){\rm Vec} (O_n,M_n,\hat{\mu}_{{\mathbf{x}}})
\\
&=& g ( [{\rm Vec} (O_n,M_n,\hat{\mu}_{{\mathbf{x}}})] )
\end{eqnarray*}
for a certain mapping $g: \mathcal{R}^{(p^2H+pH+p)} \rightarrow \mathcal{R}^{(p^2H+pH)}$
such that
$${\rm Vec}(C,M)=g( [{\rm Vec}(O,M,\mu_{{\mathbf{x}}})] ).$$
Thus the close form of $\Delta$ can be obtained by Cram\'{e}r's theorem \citep{cramer1946}:
\begin{eqnarray}
\Delta= [ \dot{g} ( [{\rm Vec}(O,M,\mu_{{\mathbf{x}}})] ) ]
    \Delta_0
    [ \dot{g} ( [{\rm Vec}(O,M,\mu_{{\mathbf{x}}})] ) ]',\label{delta}
\end{eqnarray}
where the $(p^2H+pH) \times (p^2H+pH+p)$ derivative matrix
\begin{eqnarray}
\dot{g}\left[{\rm Vec}(O,M,\mu_{{\mathbf{x}}})\right]=
\left(\begin{array}{ccc}
{\rm I}_{p^2H} & -{\rm I}_H\otimes\mu_{{\mathbf{x}}}\otimes {\rm I}_p-{\rm I}_{pH}\otimes \mu_{{\mathbf{x}}} & \dot{g}_{13}\\
0 & {\rm I}_{pH} & 0
\end{array}\right) \label{gdot}
\end{eqnarray}
with $
\dot{g}_{13} =-\left(\ldots,{\rm I}_p\otimes{\rm E}({\mathbf{x}}'|\tilde{y}=h),\ldots\right)'-{\rm Vec}(M)\otimes {\rm I}_p
$ .

In summary, to compose a consistent estimate of matrix $W$,
one can (i) substitute the usual sample moments to get the sample estimate of $\Delta_0$;
(ii) estimate $\Delta$ by substituting the usual
sample estimates for ${\rm E}({\mathbf{x}}'|\tilde{y}=h)$, $\mu_{{\mathbf{x}}}$ and $M$
in (\ref{delta}) and (\ref{gdot}); (iii) obtain the usual sample estimates of
$\Gamma_{12}$ and $\Gamma_{22}$ from the singular value decomposition of $\hat U_n$;
(iv) substitute the usual sample estimates for $F$, $G$, ${\Sigma}_{{\mathbf{x}}}$, $\Gamma_{12}$
and $\Gamma_{22}$ in Theorem \ref{theoremasavetwo} to form an estimate of $W$.
Note that both $\Delta$ and $\Delta_0$ do not rely on $\alpha$.
This fact can save a lot of computational time when multiple $\alpha$'s need to be checked.

To approximate a linear combination of chi-squared random variables,
one may use the statistic proposed by \citet*{satterthwaite1941},
\citet*{wood1989}, \citet*{satorra1994}, or \citet*{bentler2000}.
In the next applications, we will present tests based on Satterthwaite's statistic
 for illustration purpose.

\subsection{Choosing Optimal $\alpha$}
\label{sectionchoosealpha}


\citet*{ye2003} proposed a bootstrap method to pick up the ``best" linear combination of two known methods
in terms of variability of the estimated CDRS $\hat{S}_{y|{\mathbf{z}}}$.
The bootstrap method works reasonably well with known dimension $d$
of the regression,
although less variability may occur with a wrong $d$
(see Section~\ref{simulationstudy} for an example).
Another drawback is its computational intensity \citep*{liwang2007}.

Alternative criterion for ``optimal" $\alpha$ is based on the weighted chi-squared tests
developed for ${\rm SIMR}$. When multiple tests with
different $\alpha$ report the same dimension $d$, we simply pick up the $\alpha$ with
the smallest $p$-value. Given that the true dimension $d$ is detected,
the last eigenvector $\hat{\gamma}_d$ added into the estimated CDRS
with such an $\alpha$ is the most significant
one among the candidates based on different $\alpha$.
In the mean time, the other eigenvectors $\hat{\gamma}_1,\ldots, \hat{\gamma}_{d-1}$
with selected $\alpha$ tend to be more significant than other candidates too.
Based on simulation studies (Section~\ref{simulationstudy}), the performance
of the $p$-value criterion is comparable with the bootstrap one with known $d$.
The advantages of the former include that it is compatible with
the weighted chi-squared tests and it requires much less computation.

When a model or an algorithm is specified for the data analysis, cross-validation
could be used for choosing optimal $\alpha$ too, just like how people did for
model selection. For example, see \citet[chap. 7]{hastie2001}.
It will not be covered in this paper since we aim at model-free dimension reduction.

\section {Simulation Study}
\label{simulationstudy}

\subsection{A Simulated Example}
\label{simulatedexample}

Let the response $y=2z_1\epsilon + z_2^2 + z_3$,
where $({\mathbf{z}}',\epsilon)'=(z_1,z_2,z_3,z_4,\epsilon)'$
are i.i.d sample from the $N_5(0,{\rm I}_5)$ distribution.
Then the true dimension of the regression is $3$ and the true CDRS is
spanned by $(1,0,0,0)'$, $(0,1,0,0)'$, and $(0,0,1,0)'$, that is, $z_1$,
$z_2$ and $z_3$.

Theoretically,
$M_{{\rm SIR}}={\rm Diag}\{0,0,$ ${\rm Var}({\rm E}(z_3|y)),$ $0\}$,
$M_{y{{\rm -pHd}}} = {\rm Diag}\{0,2,$ $0,$ $0\}$,
and $M_{r{{\rm -pHd}}} = {\rm Diag}\left\{0,2,0,0\right\}$
have rank one and therefore are only able to
find a one-dimensional proper subspace of the CDRS.
The linear combination of any two of them suggested by \citet*{ye2003}
can at most find a two-dimensional proper subspace of the CDRS.
On the contrary,
both SAVE and SIMR are able to recover the complete CDRS at a theoretical level.

\subsection{A Single Simulation}
\label{sectionsinglesimulation}

We begin with a single simulation with sample size $n=400$.
SIR, $r$-pHd, SAVE and SIMR are applied to the data.
Number of slices $H=10$ are used for SIR, SAVE, and SIMR.
The {\tt R} package {\tt dr} \citep[version 3.0.3]{weisberg2002,weisberg2009}
is used for SIR, $r$-pHd, SAVE, as well as their corresponding marginal dimension tests.
SIMR$_\alpha$ with $\alpha=0, 0.01, 0.05$,
$0.1\sim 0.9$ paced by $0.1$, $0.95, 0.99, 1$ are applied.

For this typical simulation, SIR identifies only the direction $(.018, .000,$
$-{\mathbf{.999}},$ $-.035)'$.
It is roughly $z_3$, the linear trend.
$r$-pHd identifies only the direction $(.011,\mathbf{.999},$ $-.038,-.020)'$,
which is roughly $z_2$, the quadratic component.
As expected, SAVE works better.
It identifies $z_2$ and $z_1$.
However, the marginal dimension tests for SAVE \citep{shao2007} fail to
detect the third predictor, $z_3$.  The $p$-value of the corresponding test
is $0.331$.

Roughly speaking, SAVE with its marginal dimension test is comparable with SIMR$_{0.1}$
in this case.
The comparison between SAVE and
SIMR$_\alpha$ suggests that the failure of SAVE might due to its weights
combining the first and second inverse moments.  As $\alpha$ increases,
SIMR$_\alpha$ with $\alpha$ between $0.3$ and $0.8$ all succeed in detecting all the three
effective predictors $z_1$, $z_2$ and $z_3$.  The CDRS estimated by those candidate
matrices are similar to each other, which implies that the results with different $\alpha$
are fairly consistent.
The major difference among SIMR$_\alpha$
is that the order of the detected predictors changes roughly from
$\{z_2, z_1, z_3\}$ to $\{z_3, z_2, z_1\}$ as $\alpha$ increases from $0.3$
to $0.8$.
As expected, SIMR$_\alpha$ is comparable with SIR if $\alpha$ is close to $1$.

For this particular simulation, SIMR$_\alpha$ with $\alpha$ between $0.3$ and
$0.8$ are first selected.
If we know the true CDRS, the optimal $\alpha$
is the one minimizing the distance between the estimated CDRS and the true CDRS.
Following \citet*[p.~974]{ye2003}, the three distance
measures {\tt arccos($q$)}, $1-q$, $1-r$ behave similarly
and imply the same $\alpha=0.6$ for this particular
simulation.
Since the true CDRS is unknown, bootstrap criterion and $p$-value criterion
(Section~\ref{sectionchoosealpha}) are applied instead.

\begin{figure}[p]
\caption{Optimal $\alpha$ according to variability of $200$ bootstrapped estimated CDRS
(left panel, $d=3$ indicates the first $3$ eigenvectors considered, and so on)
or $p$-values of weighted chi-squared tests (right panel,
$d=3$ indicates the test $d\leq 2$ versus $d\geq 3$, and so on)}
\begin{center}
\hspace{0.0cm}
\psfig{figure=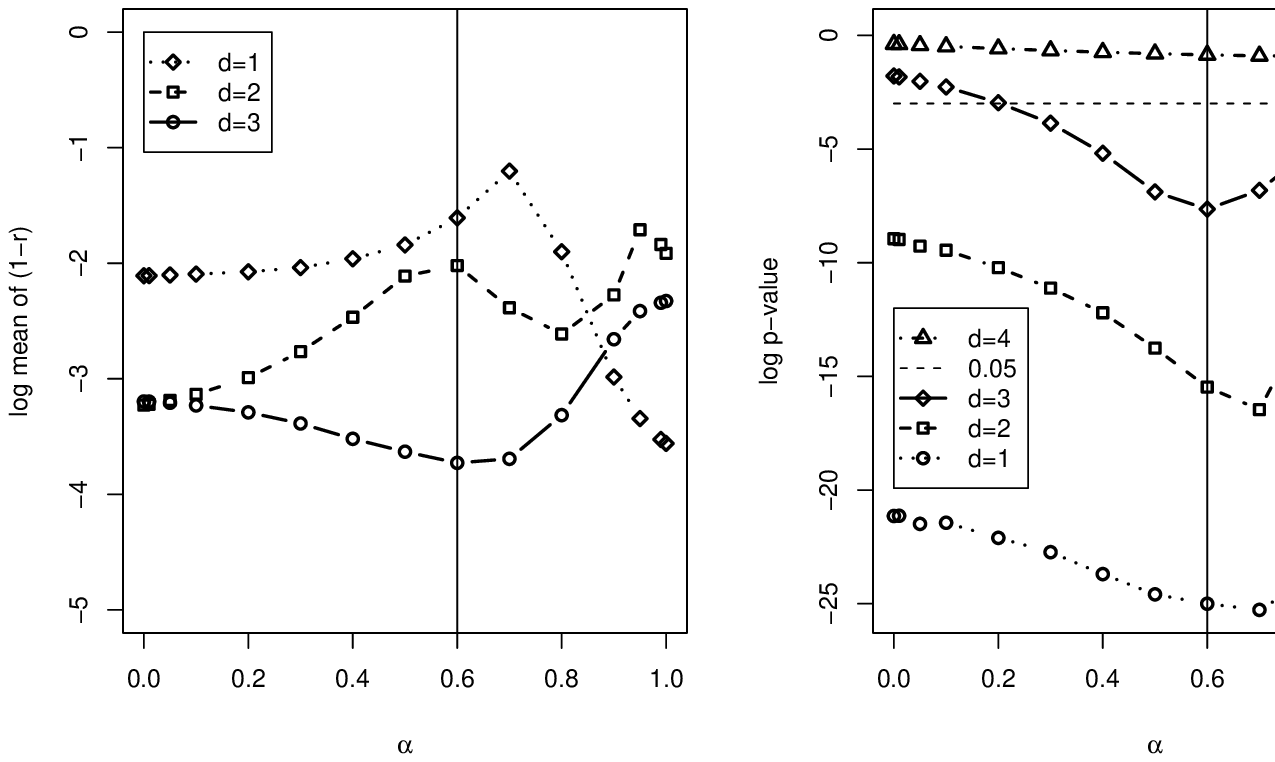,height=2.5in,width=4.5in,angle=0}
\end{center}
\label{figure12}
\end{figure}

The left panel of Figure~\ref{figure12} shows the variability of bootstrapped estimated CDRS.
Distance $1-r$ is used because it is comparable across different dimensions.
The minimum variability is attained at $d=3$ and $\alpha=0.6$, which happens to
the optimal one based on the truth.
Another 200 simulations reveal that
about $75\%$ ``optimal" $\alpha$ based on bootstrap fall in $0.5\sim 0.6$.
SIMR with $\alpha$ chosen by bootstrap
criterion attains $1-r=0.0086$ away from the true CDRS on average.
Note that low variability not necessarily implies that
the estimated CDRS is accurate.
For example, SIMR$_1$ or SIR can only detect one direction $z_3$.
However the estimated one-dimensional CDRS is fairly stable
under bootstrapping (see Figure~\ref{figure12}).

The right panel in Figure~\ref{figure12} shows that the $p$-value
criterion also picks up $\alpha=0.6$ for this single simulation
(check the line $d=3$, which is the highest one that still goes below
the significance level $0.05$).  Based on the same $200$ simulations,
about $80\%$ of the ``best" $\alpha$ selected by $p$-value criterion
fall between $0.4$ and $0.7$.  On average,
SIMR with $\alpha$ selected by $p$-values attains $1-r=0.0082$,
which is comparable with the bootstrap ones.

\subsection{Power Analysis}

We conduct 1000 independent simulations
and summarize in Table~\ref{empiricalpower05} the empirical powers and sizes of the marginal dimension tests with significance level 0.05 for
SIR, SAVE, $r$-pHd, and SIMR$_{\alpha}$ with $\alpha$ chosen by the $p$-value criterion.
For illustration purpose, we omit the simulation results of $y$-pHd
because there is little difference between $y$-pHd and $r$-pHd in this case.
The empirical powers and sizes with significance level 0.01 are omitted too
since their pattern is similar to Table~\ref{empiricalpower05}.

In Table~\ref{empiricalpower05}, the rows
$d\leq 0$, $d\leq 1$, $d\leq 2$ and $d\leq 3$ indicate different null hypotheses.
Following \citet*{bura2001}, the numerical entries in the rows $d\leq 0$,
$d\leq 1$, and $d\leq 2$ are empirical estimates of the powers of the corresponding tests,
while the entries in the row $d\leq 3$ are empirical estimates of the sizes of the tests.

As expected, SIR claims $d=1$ in most cases.
$r$-pHd works a little better.  At the significance level 0.05,
$r$-pHd has about $30\%$ chance to find out $d\geq 2$ (Table~\ref{empiricalpower05}).
At level 0.01, the chance shrinks to about $15\%$.
Both SAVE and SIMR perform much better than
SIR and pHd.   Compared with SAVE,
SIMR has consistently greater powers for the null hypotheses $d\leq 0$,
$d\leq 1$ and $d\leq 2$
across different choices of sample size, number of slices and significant level.
For example, under the null hypothesis $d\leq 2$ with sample size $400$,
the empirical powers of SIMR at
level 0.05 are $0.939$ under 5 slices and $0.943$ under 10 slices,
while the corresponding powers of SAVE are only $0.399$ and $0.213$ respectively
(Table~\ref{empiricalpower05}).
Those differences become even bigger at level 0.01.
The empirical sizes of SIMR are roughly under the nominal size 0.05
although they tend to be larger than the others.

For comparison purpose, the methods
{\it inverse regression estimator} (IRE) \citep*{cookni2005,wen2007,weisberg2009})
and {\it directional regression} (DR) \citep*{liwang2007}
are also applied. Roughly speaking, IRE performs
similar to SIR in this example.  Given that the truth dimension $d=3$ is known,
both DR and SIMR are among the best in terms of mean($1-r$).
For example, at $n=600$, DR achieves mean($1-r$) $=0.0050$ with $H=5$, $0.0053$ with $H=10$
and $0.0059$ with $H=15$, while SIMR's are $0.0048$, $0.0046$, and $0.0053$.
Nevertheless, the powers of the marginal tests for DR are between
SAVE and SIMR in this case. Roughly speaking, DR's power tests are comparable
with SIMR$_\alpha$'s with $\alpha$ between $0.2$ and $0.3$.
For example, at $H=10$ and level $0.05$,
the empirical powers of DR against $d\leq 2$
are $0.247$ with $n=200$, $0.800$ with $n=400$, and $0.974$ with $n=600$.

Among the six dimension reduction methods applied, SIMR is the most reliable one.
Besides, the chi-squared tests for SIMR do not seem to be very sensitive to the numbers of slices.
Nevertheless, we suggest that the number of slices should not be greater than 3\%-5\% of the sample size
based on the simulation results.

\section{A Real Example: Ozone Data}
\label{sectionozonedata}

To examine how SIMR works in practice, we consider a data set taken from \citet*{breiman1985}.
The response {\tt Ozone} is the daily ozone concentration in parts per million,
measured in Los Angeles basin, for 330 days in 1976.
For illustration purpose, the dependence of {\tt Ozone} on the following four predictors
is studied next:
{\tt Height}, Vandenburg 500 millibar height in meters;
{\tt Humidity} in percents;
{\tt ITemp}, Inverse base temperature in degrees Fahrenheit;
and {\tt STemp}, Sandburg Air Force Base temperature in degrees Fahrenheit.

To meet
both the linearity condition
and the constant covariance condition,
simultaneously power transformations on the predictors are estimated
to improve the normality of their joint distribution.
After replacing {\tt Humidity}, {\tt ITemp}, and {\tt STemp} with
${\tt Humidity}^{1.68}$, ${\tt ITemp}^{1.25}$, and ${\tt STemp}^{1.11}$ respectively,
SIR, $r$-pHd, SAVE and SIMR are applied to the data.
For SIR, SAVE, and SIMR, various numbers of slices are applied, and the results are
fairly consistent. Here we only present the outputs based on $H=8$.

At significance level $0.05$, SIR suggests the dimension of the regression $d=1$,
while $r$-pHd claims $d=2$.
Using the visualization tools described by \citet*{cook1994} and \citet{cook1998b},
the first pHd predictor appears to be somewhat symmetric about the response $Ozone$,
and the second pHd predictor seems to be similar to the first SIR predictor,
which are not shown in this article.
The symmetric dependency explains why SIR is not able to find the first pHd predictor.
The resulting inference based on pHd is therefore
more reliable than the inference based on SIR.

When checking the predictors of SAVE,
visual tools show a clear quadratic or even higher order polynomial dependency
between the response and the first SAVE predictor.
The second SAVE predictor is similar to the second pHd predictor,
and the third SAVE predictor is similar to the first pHd predictor.
Both SIR's and pHd's tests miss the first SAVE predictor.

Now apply SIMR to the ozone data. Bootstrap criterion picks up $\alpha=0.2$
while $p$-value criterion suggests $\alpha=0$. Nevertheless, both
SIMR$_{0.2}$ and SIMR$_0$ lead to very similar estimated CDRS in this case
(see Table~\ref{ozonetabletwo}).
As expected , they recovers all the three SAVE predictors.
Actually, those three estimated CDRS appear to be almost identical.

\section{Discussion}



SIMR$_\alpha$ and SAVE are theoretically equivalent since that the subspaces spanned by their
underlying matrices are identical.  Nevertheless, simulation study shows that
SIMR$_\alpha$ with some chosen $\alpha$ may perform better than SAVE.
The main reason is that SAVE is only a fixed combination of the first two
inverse moments.
The simulation example in Section~\ref{simulationstudy} implies that any fixed combination can not always be
the winner.
Apparently, SIMR$_{0.6}$ can not always be the winner either.  For example, if the simulation
example is changed to $y=2 z_1\varepsilon+z_2^2+0.1 z_3$, SIMR$_\alpha$ with $\alpha$ closer to $1$ will
perform better.
For practical use, multiple methods, as well as their combinations,
should be tried and unified.
SIMR$_\alpha$ with $\alpha\in (0,1)$ provide a simple solution to it.


As a conclusion, we propose SIMR using weighted chi-squared tests
as an important class of dimension reduction methods,
which should be routinely considered during the search for the central dimension reduction
subspace and its dimension.

\section*{Appendix}

{\bf Proof of Lemma \ref{asaveone}:}
By definition,
${\rm Span} \{ {\rm E} ( MM' ) \} \subseteq {\rm Span} \{ M(\omega), \omega \in \Omega_0 \} $,
if $P(\Omega_0)=1$.
On the other hand, for any $v_{p \times 1} \neq 0$,
\begin{eqnarray*}
& &v' {\rm E} ( M(\omega)M'(\omega) ) =0
\>\Rightarrow\> v' {\rm E} ( M(\omega)M'(\omega) )v =0
\\
&\Rightarrow& {\rm E} ( [v'M(\omega)] [v'M(\omega)]' ) =0
\>\Rightarrow\> [v'M(\omega)] \equiv 0, \mbox{ with probability }1
\end{eqnarray*}
Since $\{v: v'{\rm E}(MM')=0\}$ only has finite dimension, there exists an $\Omega_0$ with probability 1, such that,
$$\dim( {\rm Span} \{ {\rm E} ( M(\omega)M'(\omega) ) \} )
        \geq \dim( {\rm Span} \{ M(\omega), \omega \in \Omega_0 \} ).$$
Thus,
${\rm Span} \{ {\rm E} ( M(\omega)M'(\omega) ) \} = {\rm Span} \{ M(\omega), \omega \in \Omega_0 \}$

\vspace{0.2cm}
\noindent
{\bf Proof of Corollary \ref{asavethree}:}
\begin{eqnarray*}
{\rm Span}\{ M_{y{\rm -pHd}} = {\rm E}[ y (\mu_2(y)-{\rm I}) ] \} \subseteq {\rm Span} \{ [\mu_2(y)-{\rm I}], \forall y \} = {\rm Span} \{ M_{{\mathbf{z}}{\mathbf{z}}'|y} \}.
\end{eqnarray*}

\vspace{0.2cm}
\noindent
{\bf Proof Proposition \ref{asavefour}: }\
Define $\mu_{i}= {\rm E} [ ({\mathbf{z}}{\mathbf{z}}'-{\rm I})|y=a_i ]={\rm E} ({\mathbf{z}}{\mathbf{z}}'|y=a_i)-{\rm I}$
and $f_i={\rm Pr}(y=a_i)$ for $i=0,...k$,
then $\Sigma_{i=0}^k f_i = 1$ and $\Sigma_{i=0}^k f_i \mu_{i} = {\rm E} ( ({\mathbf{z}}{\mathbf{z}}'-{\rm I}))=0$.
The rest of the steps follow the exactly same proof as
in \citet*[A.3. Proposition 4]{yin2002}.

\vspace{0.2cm}
\noindent
{\bf Proof of Proposition \ref{asavefive}:}\
By Lemma \ref{asaveone},
\begin{eqnarray*}
{\rm Span}\{ M_{{\rm SAVE}} \} &=& {\rm Span}\{ [ \mu_1(y)\mu_1(y)' + (\mu_2(y)-{\rm I})], \forall y \}
\\
&\subseteq& {\rm Span}\{ \mu_1(y), \forall y \} + {\rm Span}\{ (\mu_2(y)-{\rm I}), \forall y \}
\\
&=& {\rm Span}\{ M_{{\rm SIR}} \} + {\rm Span}\{ M_{{\mathbf{z}}{\mathbf{z}}'|y} \}
\\
&\subseteq& {\rm Span}\{ M_{{\rm SIR}} \} + [ {\rm Span}\{ \mu_1(y)\mu_1(y)' + (\mu_2(y)-{\rm I}), \forall y \}
\\
&&   + {\rm Span}\{ \mu_1(y), \forall y \} ]
\\
&\subseteq& {\rm Span}\{  M_{{\rm SIR}} \} + {\rm Span}\{  M_{{\rm SAVE}} \} + {\rm Span}\{  M_{{\rm SIR}} \}
\\
&=& {\rm Span}\{  M_{{\rm SAVE}} \}.
\end{eqnarray*}

\vspace{0.2cm}
\noindent
{\bf Proof of Proposition \ref{asaveseven}:}\
By Lemma \ref{asaveone},
\begin{eqnarray*}
{\rm Span}\{ M^{(\alpha)}_{{\rm SIMR}} \}
&=& {\rm Span}\{ (\mu_1(y), [\mu_2(y)-{\rm I}]), \forall y \}
\\
&=& {\rm Span}\{ \mu_1(y), \forall y \} + {\rm Span}\{ [\mu_2(y)-{\rm I}], \forall y \}
\\
&=& {\rm Span}\{ M_{{\rm SIR}} \} + {\rm Span}\{ M_{{\mathbf{z}}{\mathbf{z}}'|y} \}
\\
&=& {\rm Span}\{  M_{{\rm SAVE}} \}.
\end{eqnarray*}


\vspace{0.2cm}
\noindent
{\bf Proof of Theorem~\ref{theoremasavetwo}:}\
Actually, $B= {\Sigma}_{{\mathbf{x}}}^{-1/2} (C, M) K$,
\[
\hat{U}_n
= \hat{\Sigma}_{{\mathbf{x}}}^{-1/2} (C_n, M_n)
    \left( \begin{array}{cc}
        \sqrt{1-\alpha}\ (\hat{F} \hat{G})  \otimes \hat{\Sigma}_{{\mathbf{x}}}^{-1/2} & 0
        \\
        0 & \sqrt{\alpha}\ \hat{F} \hat{G}
    \end{array}
    \right).
\]
Note that
$\left( \Gamma_{12}'B_1, \Gamma_{12}'B_2 \right)  = 0_{(p-d) \times (pH+H)}$,
$B_1 \Gamma_{221} + B_2 \Gamma_{222} = 0_{p \times (pH+H-d)}$,
${\rm Span}\{C' {\Sigma}_{{\mathbf{x}}}^{-1/2} \Gamma_{12} \} \subseteq {\rm Span}\{1_H \otimes {\rm I}_p \}$,
${\rm Span}\{M' {\Sigma}_{{\mathbf{x}}}^{-1/2} \Gamma_{12} \} \subseteq {\rm Span}\{1_H \}$,
$1'_H \hat{F}$ $=$ $0$,
$1'_H F = 0$.
Writing $\hat{{\rm I}}_p=\hat{\Sigma}_{{\mathbf{x}}}^{-1/2} \Sigma_{{\mathbf{x}}}^{1/2}$,
\begin{eqnarray*}
&&  \sqrt{n} \Gamma_{12}' \hat{U}_{n} \Gamma_{22}
\\
&=& \sqrt{n} \Gamma_{12}' \hat{U}_{n,1} \Gamma_{221}
    +\sqrt{n} \Gamma_{12}' \hat{U}_{n,2} \Gamma_{222}
\\
&=& \sqrt{1-\alpha}\ \sqrt{n} \Gamma_{12}'
        (\hat{{\rm I}}_p -{\rm I}_p +{\rm I}_p) \Sigma_{{\mathbf{x}}}^{-1/2}
        (C_n-C+C)
        [(\hat{F} \hat{G} - FG + FG) \otimes {\rm I}_p]
\\
&&
        ({\rm I}_H \otimes \Sigma_{{\mathbf{x}}}^{-1/2}) [{\rm I}_H \otimes (\hat{{\rm I}}'_p -{\rm I}_p +{\rm I}_p)]
        \Gamma_{221}\ +\ \sqrt{\alpha}\ \sqrt{n} \Gamma_{12}'
        (\hat{{\rm I}}_p -{\rm I}_p +{\rm I}_p)
\\
&&
    \Sigma_{{\mathbf{x}}}^{-1/2}
        (M_n-M+M)
        (\hat{F} \hat{G} - FG + FG)
        \Gamma_{222}
\\
&=& \sqrt{1-\alpha}\ \sqrt{n} \Gamma_{12}'
        \Sigma_{{\mathbf{x}}}^{-1/2}
        (C_n-C)
        [FG \otimes {\rm I}_p]
        ({\rm I}_H \otimes \Sigma_{{\mathbf{x}}}^{-1/2})
        \Gamma_{221}
\\
&&  +\sqrt{\alpha}\ \sqrt{n} \Gamma_{12}'
        \Sigma_{{\mathbf{x}}}^{-1/2}
        (M_n-M)
        FG
        \Gamma_{222}
   \ +\ O_p(n^{-1/2})
\\
&=& \sqrt{n} \Gamma_{12}' {\Sigma}_{{\mathbf{x}}}^{-1/2}
    [(C_n,M_n) - (C,M)] K \Gamma_{22}
   \ +\ O_p(n^{-1/2}).
\end{eqnarray*}

Therefore, the asymptotic distribution of
$\Gamma_{12}' \hat{U}_{n} \Gamma_{22}$
is determined only by
the asymptotic distribution of $(C_n,M_n)$.

\vspace{0.3cm}
\noindent
{\bf The detail of $\Delta_0$, $(p^2H+pH+p)\times (p^2H+pH+p)$:}
\[
\Delta_0=\left(
\begin{array}{ccc}
\Delta_0^{1,1} & \Delta_0^{1,2} & \Delta_0^{1,3}\\
\Delta_0^{2,1} & \Delta_0^{2,2} & \Delta_0^{2,3}\\
\Delta_0^{3,1} & \Delta_0^{3,2} & \Delta_0^{3,3}
\end{array}
\right),
\]
where
$\Delta_0^{1,1}={\rm diag}\left\{\ldots,{\rm Cov}\left({\rm Vec}({\mathbf{x}}{\mathbf{x}}')|\tilde{y}=h\right)/f_h,\ldots\right\},\> p^2H\times p^2H$;
$\Delta_0^{2,1}$ $=$ ${\rm diag}\left\{\ldots,{\rm Cov}\left({\mathbf{x}},{\rm Vec}({\mathbf{x}}{\mathbf{x}}')|\tilde{y}=h\right)/f_h,\ldots\right\},\> pH\times p^2H$;
$\Delta_0^{2,2}={\rm diag}\{$ $\ldots,$ ${\rm Cov}({\mathbf{x}}|\tilde{y}=h)/f_h,$ $\ldots\},\> pH\times pH$;
$\Delta_0^{3,1}=\left[\ldots,{\rm Cov}\left({\mathbf{x}},{\rm Vec}({\mathbf{x}}{\mathbf{x}}')|\tilde{y}=h\right),\ldots\right],$ $p\times p^2H$;
$\Delta_0^{3,2}=\left[\ldots,{\rm Cov}\left({\mathbf{x}}|\tilde{y}=h\right),\ldots\right],\> p\times pH$;
$\Delta_0^{3,3}=\Sigma_{{\mathbf{x}}},\> p\times p$;
$\Delta_0^{1,2}=\left(\Delta_0^{2,1}\right)'$;
$\Delta_0^{1,3}=\left(\Delta_0^{3,1}\right)'$;
$\Delta_0^{2,3}=\left(\Delta_0^{3,2}\right)'$.

\begin{table}[p]
\caption{Empirical Power and Size of Marginal Dimension Tests
for SIR, SAVE, SIMR$_\alpha$ with $\alpha$ Chosen by $p$-Value Criterion, and $r$-pHd,
as Well as Mean of $1-r$ Distances between Estimated 3-Dim CDRS and True CDRS, Based on 1000 Simulations
(Significance Level: 0.05; Sample Size: 200, 400, 600; Number of Slices: 5, 10, 15)}
\footnotesize
\begin{center}
\begin{tabular}{ccccccccccc}\hline\hline
\multicolumn{11}{c}{n=200}\\ \hline
 & \multicolumn{3}{c|}{SIR} & \multicolumn{3}{c|}{SAVE} &
\multicolumn{3}{c|}{SIMR$_\alpha$} & $r$-pHd\\ \cline{2-11}
Slice     & 5 & 10 & 15 & 5 & 10 & 15 & 5 & 10 & 15 & - \\ \hline
$d\leq 0$ & 0.996 & 0.967 & 0.933 & 1.000  & 0.994  & 0.885  & 1.000  &  0.999 &  0.985 & 1.000  \\ \hline
$d\leq 1$ & 0.050 & 0.053 & 0.102 & 0.561  & 0.379  & 0.152  & 0.892  &  0.855 &  0.760 & 0.277  \\ \hline
$d\leq 2$ & 0.004 & 0.003 & 0.003 & 0.061  & 0.025  & 0.007  & 0.489  &  0.441 &  0.354 & 0.027  \\ \hline
$d\leq 3$ & 0.001 & 0.000 & 0.000 & 0.003  & 0.001  & 0.000  & 0.032  &  0.022 &  0.026 & 0.005  \\ \hline
mean($1-r$) & 0.124 & 0.127 & 0.119 & 0.045 & 0.060 & 0.077  & 0.033  & 0.033  &  0.039 & 0.111 \\ \hline\hline
\multicolumn{11}{c}{n=400}\\ \hline
 & \multicolumn{3}{c|}{SIR} & \multicolumn{3}{c|}{SAVE} &
\multicolumn{3}{c|}{SIMR$_{\alpha}$} &  $r$-pHd\\ \cline{2-11}
Slice & 5 & 10 & 15 & 5 & 10 & 15 & 5 & 10 & 15 & - \\ \hline
$d\leq 0$ & 1.000 & 1.000 & 1.000 & 1.000 &  1.000 &  1.000 &  1.000 &   1.000 &   1.000 & 1.000   \\ \hline
$d\leq 1$ & 0.039 & 0.050 & 0.108 & 0.983 &  0.974 &  0.888 &  1.000 &   1.000 &   0.993 & 0.293   \\ \hline
$d\leq 2$ & 0.003 & 0.001 & 0.012 & 0.399 &  0.213 &  0.091 &  0.939 &   0.943 &   0.860 & 0.026   \\ \hline
$d\leq 3$ & 0.001 & 0.000 & 0.000 & 0.015 &  0.013 &  0.010 &  0.052 &   0.040 &   0.033 & 0.002   \\ \hline
mean($1-r$) & 0.127 & 0.129 & 0.120 & 0.016 & 0.025 & 0.038 &  0.009 &   0.009 &   0.011 & 0.109   \\ \hline \hline
\multicolumn{11}{c}{n=600}\\ \hline
 & \multicolumn{3}{c|}{SIR} & \multicolumn{3}{c|}{SAVE} &
\multicolumn{3}{c|}{SIMR$_{\alpha}$} &  $r$-pHd\\ \cline{2-11}
Slice & 5 & 10 & 15 & 5 & 10 & 15 & 5 & 10 & 15  & - \\ \hline
$d\leq 0$ & 1.000 & 1.000 & 1.000 & 1.000 &  1.000 &  1.000 &  1.000 &  1.000 &  1.000 & 1.000     \\ \hline
$d\leq 1$ & 0.054 & 0.062 & 0.053 & 1.000 &  1.000 &  0.998 &  1.000 &  1.000 &  1.000 & 0.328     \\ \hline
$d\leq 2$ & 0.001 & 0.000 & 0.002 & 0.841 &  0.601 &  0.371 &  0.996 &  1.000 &  0.992 & 0.040     \\ \hline
$d\leq 3$ & 0.001 & 0.000 & 0.000 & 0.021 &  0.019 &  0.013 &  0.048 &  0.034 &  0.031 & 0.006     \\ \hline
mean($1-r$) & 0.123 & 0.123 & 0.125 & 0.008 & 0.010 & 0.016 &  0.005 &  0.005 &  0.005 & 0.108  \\ \hline\hline
\end{tabular}
\end{center}
\label{empiricalpower05}
\normalsize
\end{table}

\begin{table}[p]
\caption{Ozone Data: Estimated CDRS by $r$-pHd, SAVE,
SIMR$_0$, and SIMR$_{0.2}$ ($H=10$ for SAVE and SIMR)}
\footnotesize
\begin{center}
\begin{tabular}{crrrrccrrrr}
\hline
\hline
\cline{2-5} \cline{8-11}
& First & Second & Third & Fourth
& & & First & Second & Third & Fourth
\\
\hline
$r{\rm -pHd}$  &   -.113 &  0.333 & ( 0.183) & (-.194) & &   ${\rm SAVE}$  &   0.635 &   0.126 &  0.096 & (-.124)\\
               &   -.049 &  0.084 & ( -.018) & (-.012) & &                 &   -.026 &   -.031 &  0.015 & (-.026)\\
               &   0.826 &  0.939 & ( -.642) & (-.030) & &                 &   -.665 &   -.621 &  -.664 & (-.143)\\
               &   -.551 &  -.031 & ( 0.745) & (0.981) & &                 &   -.392 &   -.773 &  0.741 & (0.981)\\
\hline
${\rm SIMR}_{0}$    &   0.652 &   0.169 &  0.092 & (0.125) & &   ${\rm SIMR}_{0.2}$  &   0.685 &   0.204 &   0.092 & (-.125)\\
                    &   -.025 &   -.032 &  0.015 & (0.026) & &                       &   -.024 &   -.031 &   0.015 & (-.026)\\
                    &   -.662 &   -.803 &  -.645 & (0.137) & &                       &   -.653 &   -.708 &   -.653 & (-.141)\\
                    &   -.369 &   -.571 &  0.758 & (-.982) & &                       &   -.322 &   -.676 &   0.751 & (0.982)\\
\hline
\hline
\end{tabular}
\end{center}
Note: ``($\cdot$)" indicates nonsignificant direction at level $0.05$.
\label{ozonetabletwo}
\normalsize
\end{table}

\end{document}